\documentclass[final,5p,times,twocolumn]{elsarticle}

\usepackage{amssymb}
\usepackage{amsmath}
\usepackage{amsfonts}
\usepackage{amsthm}
\usepackage{xcolor}
\usepackage{graphicx}
\usepackage{float}
\usepackage{color,soul}

\usepackage{hyperref}
\hypersetup{
    colorlinks=true, 
    linktoc=all,     
    citecolor=black,
    filecolor=black,
    linkcolor=black,
    urlcolor=black
}

\DeclareUnicodeCharacter{2060}{}
\DeclareUnicodeCharacter{2212}{}

\biboptions{comma,compress,square}

\newcounter{bla}

\journal{Computer Physics Communications}

\begin{document}

\begin{frontmatter}

\title{RI$-$Calc: A User Friendly Software and Web Server for Refractive Index Calculation}

\author[a]{Leandro Benatto\corref{author}$^{\dagger}$}
\author[a]{Omar Mesquita}
\author[b]{Lucimara S. Roman}
\author[b]{Marlus Koehler}
\author[a,c]{Rodrigo B. Capaz}
\author[a]{Grazi\^{a}ni Candiotto\corref{author}$^{\dagger}$}

\address[a]{Institute of  Physics, Federal University of Rio de Janeiro, 21941$-$909, Rio de Janeiro$-$RJ, Brazil.}
\address[b]{Department of Physics, Federal University of Paran\'{a}, 81531$-$980, Curitiba$-$PR, Brazil.}
\address[c]{Brazilian Nanotechnology National Laboratory (LNNano), Brazilian Center for Research in Energy and Materials (CNPEM), 13083$-$100, Campinas$-$SP, Brazil.}

\cortext[author]{Corresponding authors: \\ lb08@fisica.ufpr.br and gcandiotto@iq.ufrj.br\\$\dagger$ These authors contributed equally to this work}

\begin{abstract}
The refractive index of an optical medium is essential for studying a variety of physical phenomena. One useful method for determining the refractive index of scalar materials (\textit{i.e}, materials which are characterized by a scalar dielectric function) is to employ the Kramers$-$Kronig (K$-$K) relations. The K$-$K method is particularly useful in cases where ellipsometric measurements are unavailable, a situation that frequently occurs in many laboratories. Although some packages can perform this calculation, they usually lack a graphical interface and are complex to implement and use. Those deficiencies inhibits their utilization by a plethora of researchers unfamiliar with programming languages. To address the aforementioned gap, we have developed the Refractive Index Calculator (RI$-$Calc) program that provides an intuitive and user$-$friendly interface. The RI$-$Calc program allows users to input the absorption coefficient spectrum and then easily calculate the complex refractive index and the complex relative permittivity of a broad range of thin films, including of molecules, polymers, blends, and perovskites. The program has been thoroughly tested, taking into account the Lorentz oscillator model and experimental data from a materials' refractive index database, demonstrating consistent outcomes. It is compatible with {\textit{Windows}}, {\textit{Unix}}, and {\textit{macOS}} operating systems. You can download the \href{https://github.com/NanoCalc/RICalc}{RI$-$Calc} binaries from our \textit{GitHub} repository  or conveniently access the program through our dedicated web server at \href{https://nanocalc.org/}{nanocalc.org}.
\end{abstract}


\end{frontmatter}


\section{Introduction}

The complex index of refraction (or optical constants, $\tilde{n})$ is an important property
that characterizes the interaction of light with matter\cite{singh2002refractive}. $\tilde{n}$ is formed by a real ($\eta$) and an imaginary ($\kappa$) parts that are both functions of the  wavelength ($\lambda$) so that $\tilde{n}(\lambda)=\eta(\lambda)+i\kappa(\lambda$). $\eta(\lambda)$ is the ratio of the speed of light in a vacuum compared to the phase velocity of light in the material, whereas $\kappa(\lambda)$ is the light extinction (or attenuation) coefficient due to absorption.

Several important optical and optoelectronic  phenomena  are decisively  influenced by $\tilde{n}(\lambda)$ \cite{liu2009high,khan2021refractive}. For example, $\eta(\lambda)$ is essential to calculate the F\"{o}rster resonance energy transfer (FRET) rate between molecules that are fluorescent \cite{hildebrandt2013,mikhnenko2015exciton,souza2022}⁠. The FRET process is important for photosynthesis \cite{clegg2009forster,candiotto2017} and many technological applications like biosensing \cite{chou2015forster,tian2021fluorescent}, organic light$-$emitting diodes \cite{heimel2018unicolored,candiotto2020,barreto2023}, and organic photovoltaics \cite{benatto2021conditions,du2020delayed,karki2020unifying,candiotto2024exploring}, among others \cite{alhadrami2021peptide,andre2021,sharma2021recent,merces2021,candiotto2024}. Another application example that relies heavily on $\tilde{n}(\lambda$) is the multilayer optical structure. The matrix method is employed in such devices to compute the reflection and transmission coefficients of the electromagnetic field,⁠ where $\tilde{n}(\lambda)$ of each layer is necessary for the calculations \cite{pettersson1999modeling,peumans2003small}. From this method it is possible to obtain the optical electric field distribution inside the devices, the fraction of light absorbed per layer, the external quantum efficiency, among other relevant information.

Considering the progress in materials design to improve device efficiencies \cite{zhang2022high,sun2022high,li2021asymmetric,labanti2021selenium, sai2020designing,florindo2023},⁠ a quick and efficient way to obtain the refractive index of different materials is essential and very welcomed. In many situations the ellipsometry apparatus (that can be used to measure $\tilde{n}(\lambda)$) is not easily available so that the alternative is to employ theoretical approaches. One convenient theoretical method is the application of the Kramers$-$Kronig (K$-$K) relation \cite{lucarini2005kramers}. 

There are some packages that apply the K$-$K coefficients to calculate $\tilde{n}(\lambda)$. Some of them are not user friendly and demand some knowledge of code development to be useful \cite{KKcalc}. Others of them that have a graphical interface (only for \textit{Windows}) also have many integrated features and fitting options that make their use non$-$trivial \cite{kuzmenko2005kramers,RefFit}. In addition, it is worth noting that there are instances in which utilizing the software incurs a usage fee \cite{TFCalc}. It is important to mention that some programs employ alternative approaches to compute the refractive index, one of which is the envelope method \cite{jena2021prisa}. 

This context motivated us to developed a software, Refractive Index Calculator (\href{https://github.com/NanoCalc/RICalc}{RI$-$Calc}) to evaluate $\tilde{n}(\lambda)$ using the K$-$K method. RI$-$Calc was developed in Python programming language. Compared to other software, it features a simple and intuitive graphic interface, making it very easy to handle. With a focus on investigating novel materials, RI$-$Calc allows users to input the absorption coefficient spectrum and then easily calculate the complex refractive index and the complex relative permittivity of a broad range of thin films of scalar marterials (\textit{i.e}, materials which are characterized by a scalar dielectric function) such as polymers, blends, and perovskites. The RI$-$Calc output files can be readily utilized as input for other softwares that model the FRET process \cite{benatto2023fret,benatto2024plq} or multilayer devices\cite{SETFOS}. 

\section{Method}

Before describing the capabilities of the RI$-$Calc code, it is important to review the theory behind the use of K$-$K relations to calculate $\tilde{n}(\lambda)$. As mentioned above, the complex refractive index data is traditionally obtained from spectroscopic ellipsometry \cite{van2017optical,xia2019optical,ball2015optical}. However, $\eta(\lambda)$ can also be estimated by using the K$-$K method (also known as the Hilbert transform) if the value of $\kappa(\lambda)$ can independently be evaluated using another experimental technique \cite{benatto2021conditions,kerremans2020optical,karuthedath2021intrinsic}⁠.  One possibility to obtain $\kappa(\lambda)$ is by measuring the Napierian (base$-e$) absorption coefficient spectrum $\alpha(\lambda)$ and applying it to the following equation \cite{sai2020designing,yang2008optical}

\begin{equation}\label{eq-kappa}
\kappa(\lambda)=\dfrac{\lambda \alpha(\lambda)}{4 \pi}.
\end{equation}

\noindent Care must be taken regarding the absorption coefficient. In the case of thin films, $\alpha(\lambda)$ must be obtained through methods that take into account interference effect, otherwise this can lead to extensive errors with regard to the determination of the $\kappa(\lambda)$ \cite{mayerhofer2020removing,cesaria2012realistic} It is worth mentioning that there is widespread use of the decadic (base$-$10) form of the absorption coefficient in the literature, making its specification frequently overlooked. The relationship between the Napierian absorption coefficient, $\alpha(\lambda)$, and the decadic absorption coefficient, $a(\lambda)$, is given by \cite{verhoeven1996glossary}: 

\begin{equation}\label{eq-a}
\alpha(\lambda)=a(\lambda)ln(10).
\end{equation}

\noindent With $\kappa(\lambda)$ readily available, the K$-$K relation provides a straightforward method for determining $\eta(\lambda)$. In terms of the optical wavelength, $\eta(\lambda)$ is defined as follows \cite{balawi2020quantification}

\begin{equation}\label{eq-eta}
{\eta(\lambda_{i}) = \eta(\infty) + \dfrac{2}{\pi} PV \int^{\infty}_{0} \dfrac{\lambda \kappa(\lambda) d \lambda}{\lambda^{2}-\lambda_{i}^{2}},}
\end{equation}

\noindent where $\eta(\infty)=1$ and PV denotes the Cauchy principal value. To obtain $\eta(\lambda)$ with reasonable precision it is important to know the value of $\kappa(\lambda)$ for the largest range of wavelengths possible. Due to experimental limitations, however, $\kappa(\lambda)$ will be known just for a limited interval of the spectrum. Hence $\kappa(\lambda)$ will be unknown for a relevant part of the integral range in Eq. \ref{eq-eta}. In order to overcome this obstacle, the K$-$K relation can be rewritten in the form (see the demonstration in ref.\cite{nitsche2004determination})

\begin{equation}\label{eq-eta-2}
{\eta(\lambda_{i}) = \eta_{offset} + \dfrac{2}{\pi} PV \int^{\lambda_{U}}_{\lambda_{L}} \dfrac{\lambda \kappa(\lambda) d \lambda}{\lambda^{2}-\lambda_{i}^{2}}, \lambda_{L} \leq \lambda_{i} \leq \lambda_{U},}
\end{equation}  

\noindent where $\lambda_{L}$ and $\lambda_{U}$ are the lower and upper bounds. To minimize the errors in the calculation of $\eta(\lambda)$, the constant boundary extension procedure can be applied \cite{mayerhofer2022infrared}, so that $\alpha(\lambda) = \alpha(\lambda_{L})$ for $\lambda \leq \lambda_{L}$ and $\alpha(\lambda) = \alpha(\lambda_{U})$ for $\lambda \geq \lambda_{U}$. Note that the index of refraction was shifted by a constant value, where $\eta_{offset} \geq 1$.  In relation to $\eta_{offset}$,  it is typically found to closely resemble the refractive index at the point of minimum dispersion, or a point in close proximity within the transparency region between the infrared and UV/Vis spectral range \cite{mayerhofer2022infrared}. In the transparency region $\kappa$ is negligible and $\eta_{offset}=\sqrt{\varepsilon_{r}}$, where in this specific circumstance $\varepsilon_{r}$ is often called relative dielectric constant of the medium \cite{balawi2020quantification,chandran2017direct}. Note that $\varepsilon_{r}(\lambda)$ is the complex relative permittivity of the medium (complex dielectric function), $\varepsilon_{r} = \varepsilon_{r}' - i\varepsilon_{r}''$, where \cite{whatmore2017springer,green2015optical}

\begin{equation}\label{eq-varepsilon1}{\varepsilon_{r}' = \eta^{2} - \kappa^{2}}
\end{equation}

\noindent and

\begin{equation}\label{eq-varepsilon2}{\varepsilon_{r}'' = 2\eta\kappa.}
\end{equation}  

\begin{figure}[t!]
\centering
\includegraphics[width=\linewidth]{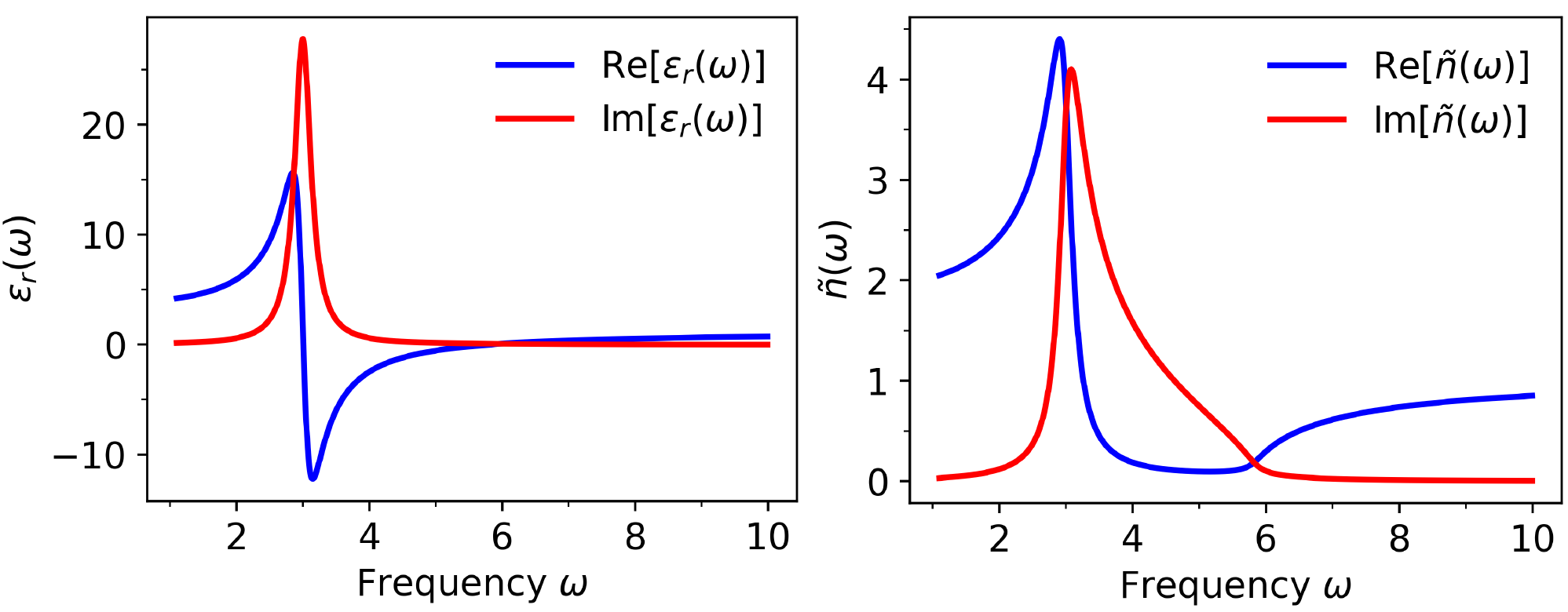}
\caption{Left) Real and imaginary parts of the complex relative permittivity ($\varepsilon_{r}$) for a single Lorentz oscillator dielectric, considering $\omega_{p}=5$, $\omega_{0} =3$ and $\Gamma=0.3$. Right) Real and imaginary parts of the complex refractive index ($\tilde{n}$) calculated from $\varepsilon_{r}$ (see, Eq. 7).}
\label{fig-lorentz-1}
\end{figure}

\begin{figure}[t!]
\centering
\includegraphics[width=\linewidth]{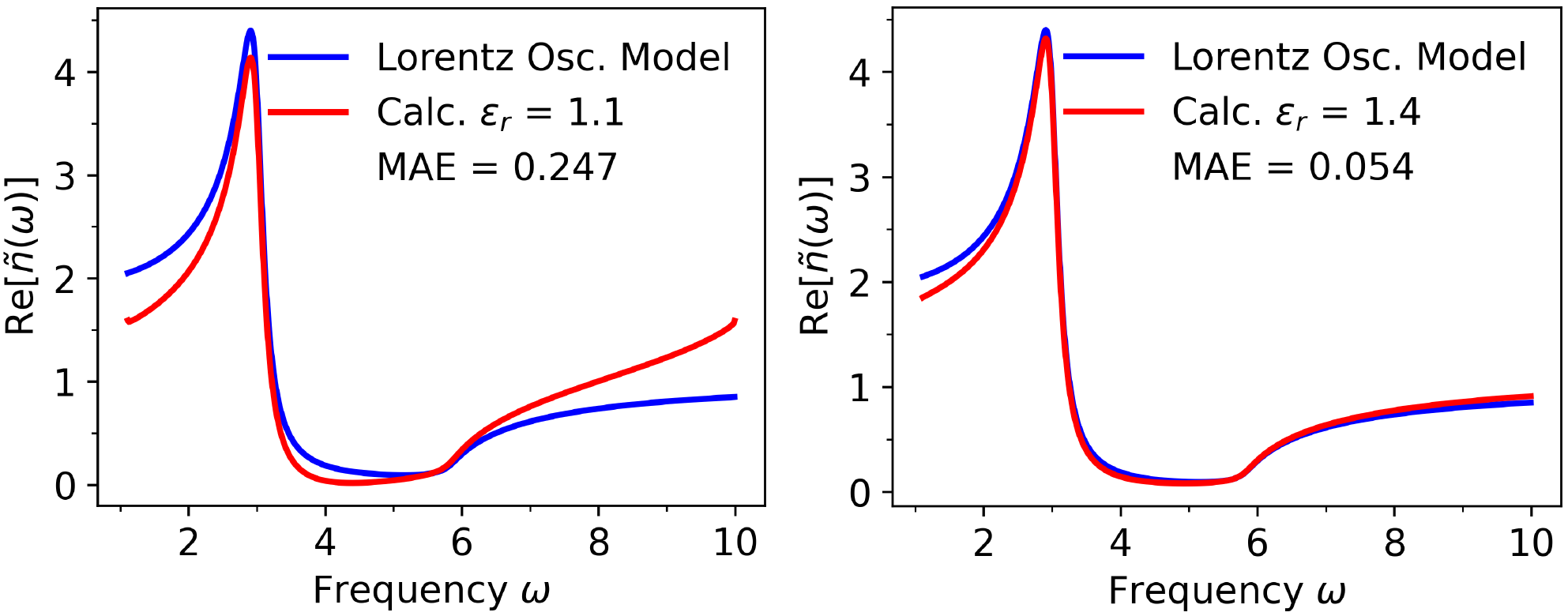}
\caption{Comparison between the real refractive index calculated by the Lorentz oscillator model and by the algorithm implemented in RI$-$Calc using $\kappa(\omega)$ as input. Left) Not applying the constant boundary extension procedure. Right) Applying the constant boundary extension procedure. To better describe the model the dielectric constant was adjusted. MAE is the mean absolute error.}
\label{fig-lorentz-2}
\end{figure}

\begin{figure*}[t!]
\centering
\includegraphics[width=\linewidth]{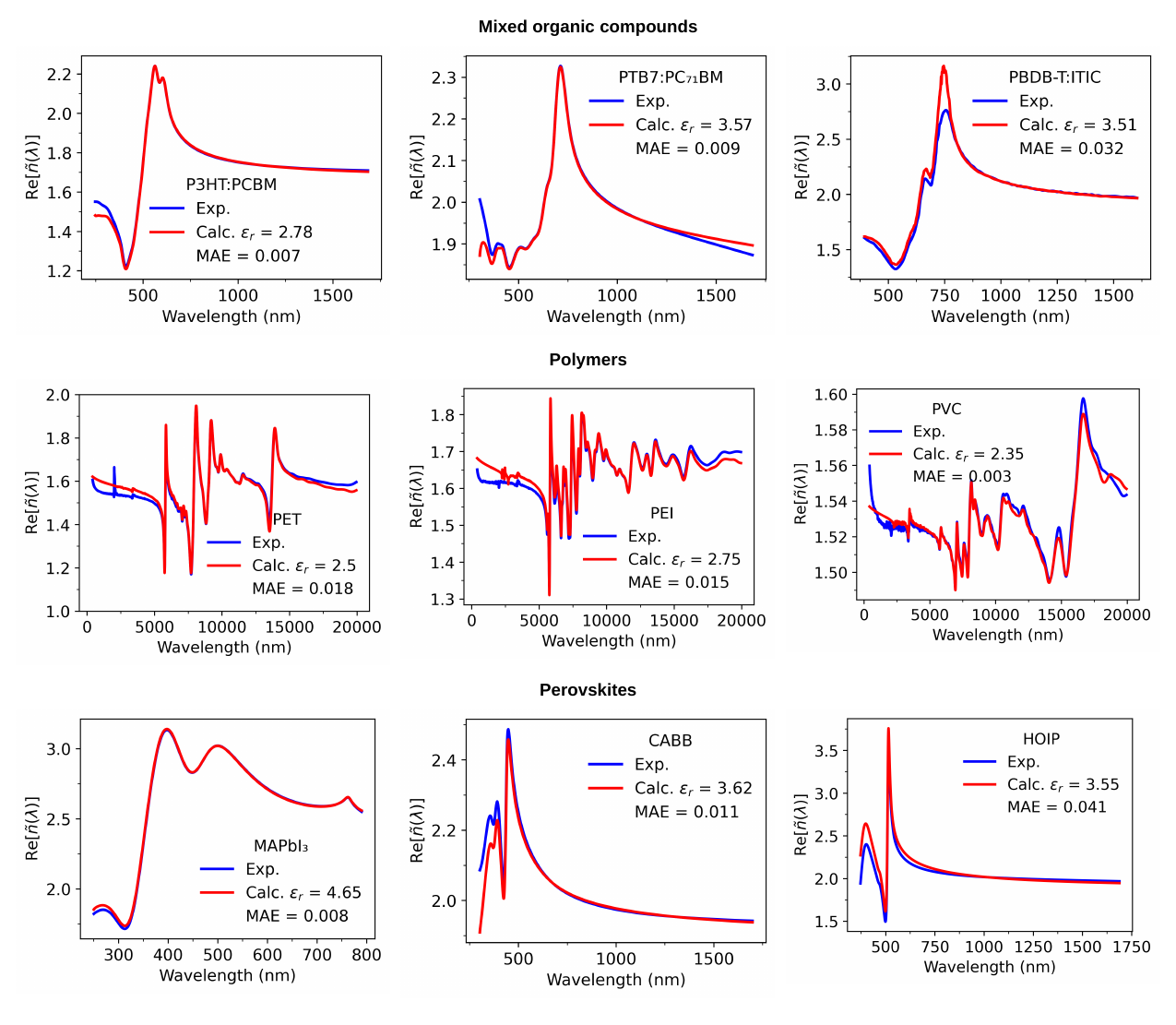}
\caption{Real part of the complex refractive index calculated from $\kappa(\lambda)$ taken from ref. \cite{refractiveindex} and its comparison with the experimental result. To better describe the experiment the dielectric constant was adjusted. MAE is the mean absolute error.}
\label{fig-theory-experiment}
\end{figure*}

Before describing the basic features of the RI$-$Calc program, we will show some tests of the algorithm developed for our software to calculate Eq. \ref{eq-eta-2} using the offset procedure combined with the constant boundary extension procedure. Firstly, the methodology used in RI$-$Calc is compared with Lorentz oscillator model, that is a classical (and analytical) model used to describe the behavior of dielectric materials in the presence of an electric field. It provides a simple yet insightful representation of how bound charges in a material respond to an external electric field, leading to the material's dielectric behavior. The model assumes that the electrons in a dielectric material are bound to their respective atoms or molecules and can be approximated as driven damped harmonic oscillators. Each oscillator represents the response of a group of bound charges to the applied electric field. Thus, the complex dielectric function $\varepsilon_{r}$ of the material can be described using the Lorentz oscillator model for multiples oscillators as follows

\begin{equation}\label{eq:lorentz_ful_sum}
\varepsilon_{r}(\omega)=1+\sum\limits_{n}\dfrac{\omega_{p,n}^{2}}{\omega_{0,n}^{2}-\omega^{2}+i\omega\Gamma_{n}},  
\end{equation}

\noindent where $\omega$, $\omega_{p,n}$, $\omega_{0,n}$, $\Gamma_{n}$ are respectively the frequency, the plasma frequency, the peak frequency, and the damping factor of the oscillator, respectively, for the $n^{th}$ type of electron\cite{liu2005characterization,ren1991ellipsometric}. For a single electron, the Eq. \ref{eq:lorentz_ful_sum} becomes

\begin{equation}\label{eq:lorentz_ful}
{\varepsilon_{r}(\omega)=1+\dfrac{\omega_{p}^{2}}{\omega_{0}^{2}-\omega^{2}+i\omega\Gamma},  }  
\end{equation}

Splitting  Eq. \ref{eq:lorentz_ful} in Re$[\varepsilon_{r}(\omega)]=\varepsilon_{r}^{\prime}(\omega)$ and Im$[\varepsilon_{r}(\omega)]=\varepsilon_{r}^{\prime\prime}(\omega)$ it is obtained

\begin{equation}\label{eq:lorentz_Re}
{\varepsilon_{r}^{\prime}(\omega)=1+\omega_{p}^{2}\left[\dfrac{\omega_{0}^{2}-\omega^{2}}{(\omega_{0}^{2}-\omega^{2})^{2}+\omega^{2}\Gamma}\right],}
\end{equation}

\noindent and

\begin{equation}\label{eq:lorentz_Re}
{\varepsilon_{r}^{\prime}(\omega)=\omega_{p}^{2}\left[\dfrac{\omega\Gamma}{(\omega_{0}^{2}-\omega^{2})^{2}+\omega^{2}\Gamma}\right],}
\end{equation}

\noindent Thus, the real and imaginary parts of the complex index of refraction can be calculated directly from the real and imaginary parts of the complex dielectric function

\begin{equation}\label{eq:etta_new}
{\eta (\omega) = \sqrt{\dfrac{\vert\varepsilon_{r}\vert+\varepsilon_{r}^{\prime}}{2}},}
\end{equation}

\noindent and

\begin{equation}\label{eq:kappa_new}
{\kappa(\omega) = \sqrt{\dfrac{\vert\varepsilon_{r}\vert-\varepsilon_{r}^{\prime}}{2}}.}
\end{equation}

\noindent Figure \ref{fig-lorentz-1} (left) illustrates the complex dielectric function, which was computed using the analytical Lorentz oscillator model. Subsequently, the real and imaginary components of the complex refractive index were derived from this calculation employing Eq. \ref{eq:etta_new} and \ref{eq:kappa_new}, respectively. These results are shown in Figure \ref{fig-lorentz-1} (right). To further evaluate the accuracy of RI$-$Calc algorithm, we focused on the imaginary part of the refractive index. Our objective was to calculate its real counterpart and subsequently compare it with the corresponding value obtained from the Lorentz oscillator model.

Figure \ref{fig-lorentz-2} (left) presents the outcome of our test without the application of the constant boundary extension procedure. As anticipated, significant deviations were observed, particularly at the spectral edges, leading to a relatively high mean absolute error (MAE). However, in Figure \ref{fig-lorentz-2} (right), we demonstrate the results of our test after applying the constant boundary extension procedure. Remarkably, this procedure substantially improved the correlation between the curves, resulting in a significantly lower MAE. Therefore, the application of the constant boundary extension procedure proved to be effective in enhancing the accuracy of our algorithm, particularly in reducing discrepancies near the spectral edges. These findings highlight the importance of such procedures in refining and validating computational models.

Once the methodology was chosen, additional tests were carried out. The tests were performed for mixed organic compounds, polymers, and perovskites employing data taken from ref. \cite{refractiveindex}, which is a database of materials' refractive index. The class of materials under consideration in this test serves as the primary focus of the RI$-$Calc application. Taking $\kappa(\lambda)$ from the database, we calculate the real part of the complex refractive index. Subsequently, we compared our calculated results with its correspondent available in the database. According to the data presented in Figure \ref{fig-theory-experiment} there is a low MAE between the $\eta(\lambda)$ estimated by the program and the database values. Therefore, choosing a meaningful offset, the $\eta(\lambda)$ calculated from Eq. \ref{eq-eta-2} is in reasonable agreement with ellipsometry measurements. The good correlation between theory and experiment demonstrates the method's consistency when applied to these classes of materials.
We would like to emphasize that, in this section, some implicit assumptions have been made, which include the absence of anisotropy in the medium. Therefore, the approach is valid for materials with low anisotropy to justify the assumption of a scalar dielectric function.\cite{mayerhofer2022infrared,mayerhofer2020bouguer} Polycrystalline samples can be analyzed under the above assumptions as long as the crystallites are randomly oriented and the crystallites are small compared to the resolution limit of the light (\textit{i.e}, the crystallite size is much smaller than the wavelength).\cite{mayerhofer2005optical}

\begin{figure}[!t]
\centering
    \includegraphics[width=\linewidth]{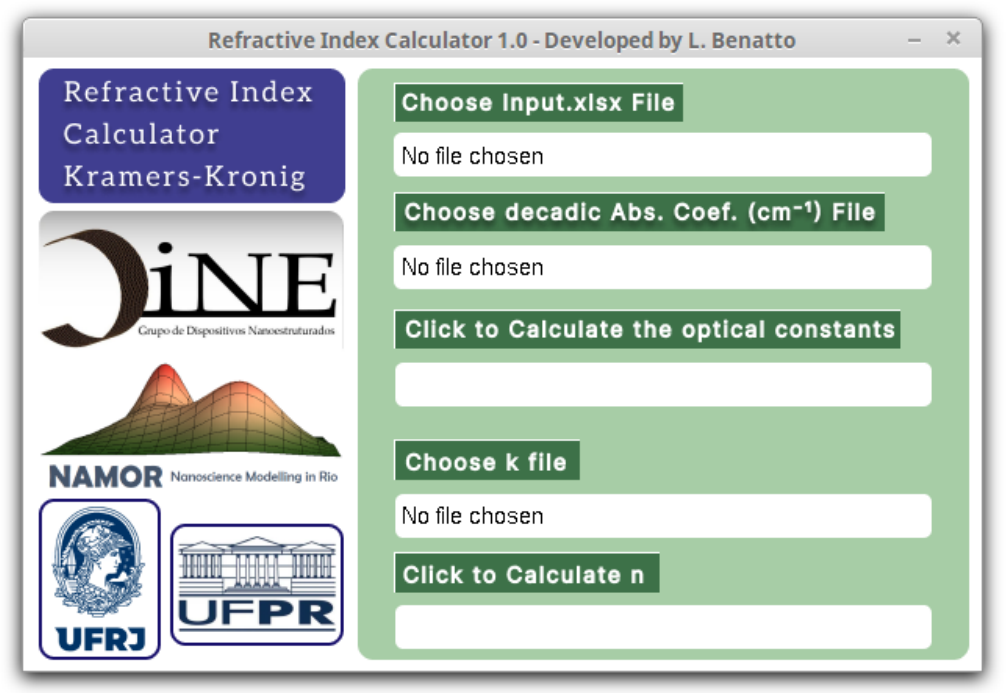}
    \caption{RI$-$Calc interface.}
    \label{fig-panel-interface}
\end{figure}

\section{Software architecture, implementation and requirements}

RI$-$Calc is a license free code and can be used \textit{via} a dedicated web server or downloading the binary files. The web server can be accessed in \href{https://nanocalc.org/}{nanocalc.org} and it is compatible with the main browsers such as \textit{Chrome}, \textit{Firefox}, \textit{Safari}, \textit{Opera}, \textit{Brave}, \textit{Edge} and \textit{Internet Explorer}. The binary files for {\textit{Unix}}, {\textit{Windows}}, and {\textit{macOS}} operational systems are available for download at \href{https://github.com/NanoCalc/RICalc}{RI$-$Calc} repository on \textit{GitHub}. The program is implemented in Python 3 (v. 3.6) \cite{van2009} and makes use of four Python libraries, namely \textit{Pandas} \cite{pandas2010}, \textit{NumPy} \cite{harris2020}⁠ and \textit{SciPy} \cite{virtanen2020}⁠ for data manipulation and \textit{Matplotlib} \cite{hunter2007}⁠ for data visualization. Specifically, the Scipy submodule scipy.fftpack.hilbert was utilized to perform the Hilbert transformation of a periodic sequence $k(\lambda)$ and solve Eq. \ref{eq-eta-2}. The software was designed to be of simple operation and take up little disk space, around 75MB.

\section{Program and Application}

Figure \ref{fig-panel-interface} shows the initial graphical interface of the RI$-$Calc software. This page requests input information on the material to be analyzed. The initial request is for an  XLSX file (easily editable in Microsoft Excel, Apple Numbers, and LibreOffice Calc, among others) containing the material name, its corresponding relative dielectric constant, and spectral scale, which may be in Wavelength (nm), Wavenumber ($cm^{-1}$) or Energy (eV). To facilitate the completion of the necessary information, an illustrative input example has been made available for download. This resource aims to guide users through the process of filling out the required fields accurately.

The next input line asks for the material's decadic absorption coefficient spectrum in units of $cm^{-1}$ (within the program Eq. \ref{eq-a} is used to obtain the Napierian absorption coefficient). At this point, we emphasize that especially for thin films with thicknesses on the order of the wavelength of the light, interference effects can not be neglected to determine the absorption coefficient. Therefore, a model that considers this effect into account is essential. Among the various possible methods to obtain the absorption coefficient, a simple procedure that considers interference effects, developed by Cesaria \textit{et al.} \cite{cesaria2012realistic}\, can be used from measured transmittance (T) and reflectanc (R) spectra of thin films. This is interesting because the experimental equipment necessary to measure the T and R spectra is much simpler compared to the laboratory apparatus needed to perform ellipsometry experiments. For further discussions about this topic, an overview of spectroscopic data interpretation, both qualitative and quantitative, is provided in ref. \cite{mayerhofer2020bouguer}.

\begin{figure}[!t]
\centering
    \includegraphics[width=\linewidth]{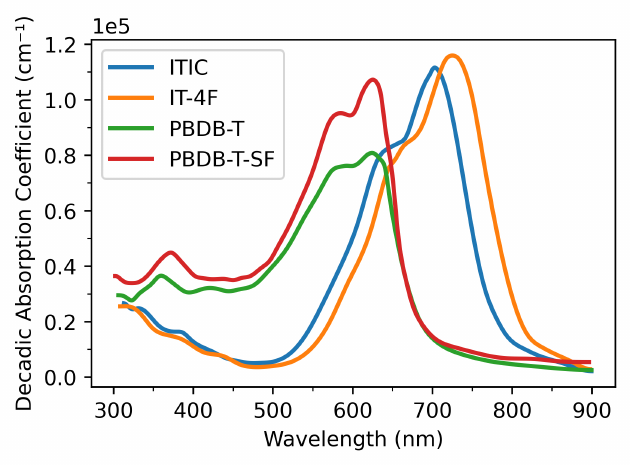}
    \caption{Decadic absorption coefficient of solid thin film of the ITIC, IT-4F, PBDB-T and PBDB-T-SF organic semiconductors. Experimental results of ref. \cite{zhao2017molecular}.}
    \label{fig-abs-coef}
\end{figure}

The spectral scale can be formatted using arbitrary steps since the program will perform a cubic interpolation of the data. Once the decadic absorption coefficient spectrum is supplied to the program, the complex refractive index and the complex dielectric function can be calculated. In case  the imaginary part of the refractive index was already known, the program has the option to calculate only the real part, see Figure \ref{fig-panel-interface}. The resulting data is printed in a ASCII file, and a graph is automatically created from the output.

We will exemplify the RI$-$Calc functionalities by taking as input the decadic absorption coefficient extracted from Figure \ref{fig-abs-coef} of Zhao \textit{et al}. \cite{zhao2017molecular}. The materials considered in this work were two molecular acceptors (ITIC and IT$-$4F) and two polymers (PBDB$-$T and PBDB$-$T$-$SF). These organic semiconductors are widely employed in organic solar cells \cite{armin2021history,zheng2020pbdb}.⁠

\begin{figure}[!t]
\centering
    \includegraphics[width=\linewidth]{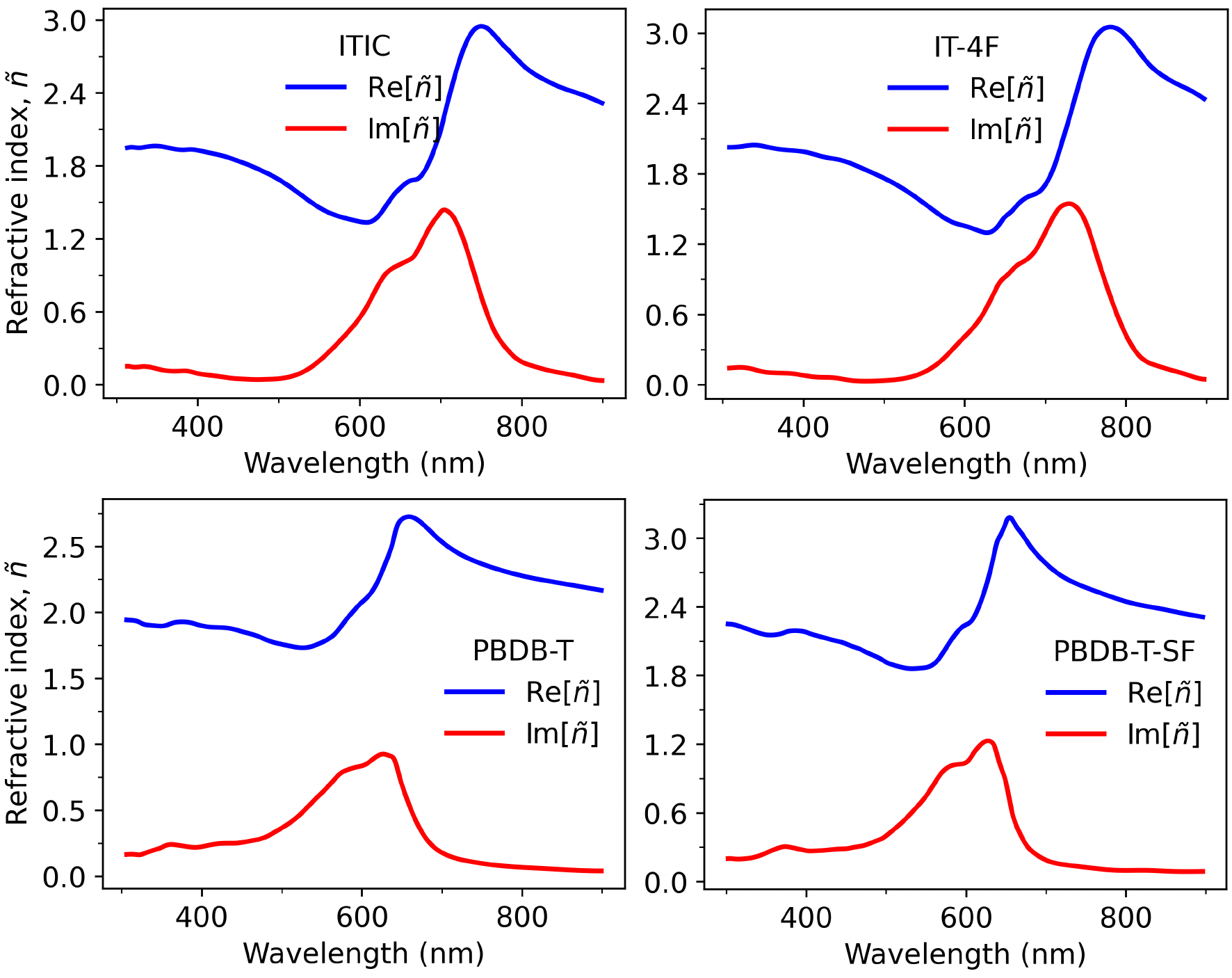}
    \caption{Complex index of refraction calculated from absorption coefficient.}
    \label{fig-nk}
\end{figure}

\begin{figure}[!t]
\centering
\includegraphics[width=\linewidth]{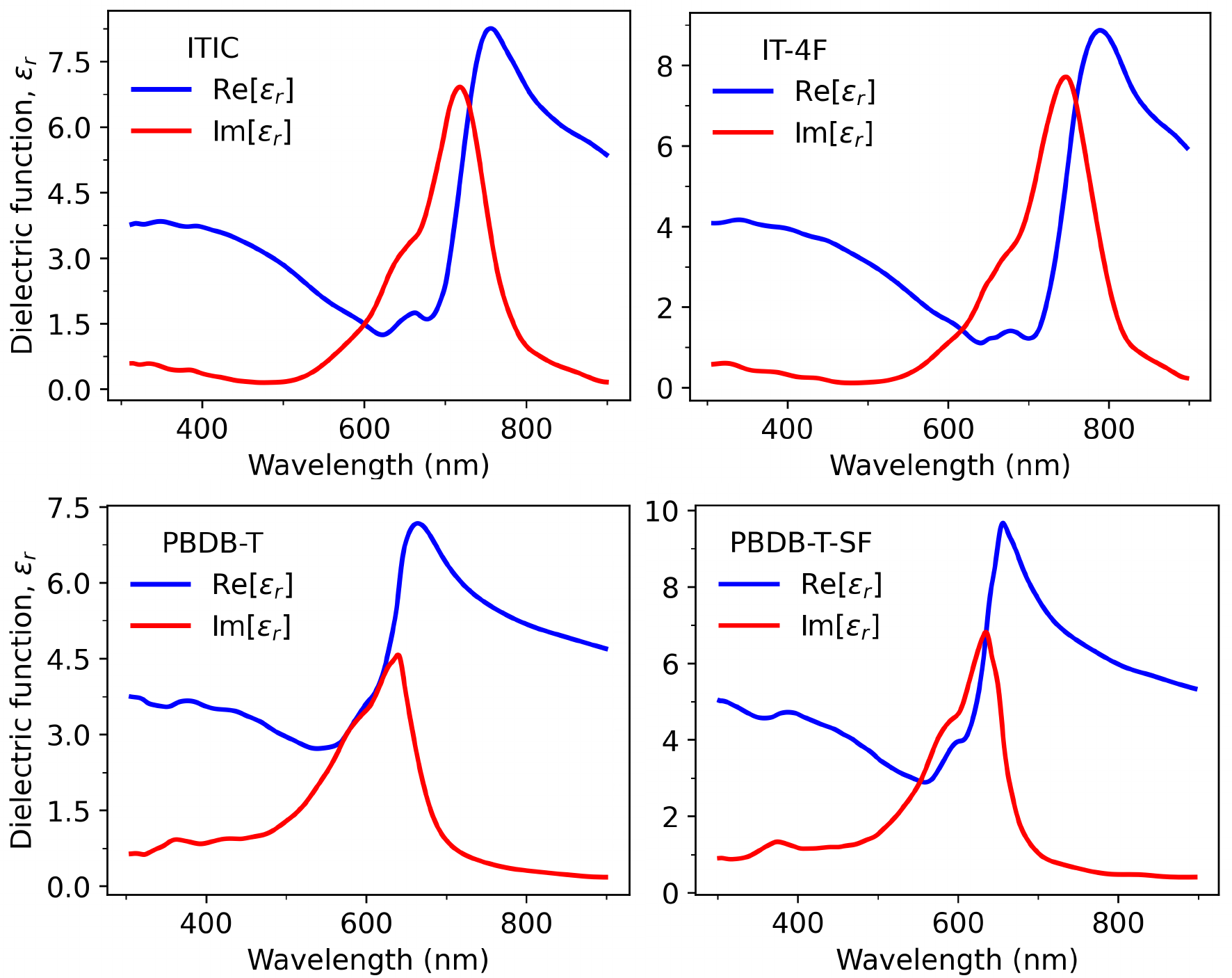}
    \caption{Complex dielectric function calculated from absorption coefficient.}
    \label{fig-e}
\end{figure}

After processing the decadic absorption coefficient spectra measured for the materials in  ref. \cite{zhao2017molecular}, RI$-$Calc calculated their complex refractive index and complex dielectric function. These results can be seen in Figure \ref{fig-nk} and Figure \ref{fig-e}. For these calculations was used the same dielectric constant of 3.5 for the four organic semiconductors. The complex refractive index found using RI$-$Calc is in good accordance with the experimental results of literature measured by the ellipsometry technique \cite{kerremans2020optical,harillo2020efficient}. This outcome provides additional evidence supporting the reliability of the calculation method.

It is worth to mention that the complex refractive index of a thin film is influenced by the production process such as the chosen solvent$/$solvent$-$additive, deposition technique, and post$-$treatment, which can impact in the film morphology \cite{cheng2014comparison,lou2011effects}. Hence some variations in $\tilde{n}(\lambda)$ are always expected for a given material due to these circumstantial influences. Therefore, it is crucial for researchers to have access to a straightforward and effective method for characterizing their samples. 

\section{Conclusions}
In summary, the theory behind RI$-$Calc along with tests of its functionality and applicability has been demonstrated in detail. The methodology employed in RI$-$Calc was compared with both the Lorentz oscillator model and experimental results from a material refractive index database. The positive outcomes from these tests clearly demonstrated the method's consistency, thereby enabling its application to a specific class of materials with low anisotropy. Therefore, the program provides the complex refractive index and the complex relative permittivity of molecular acceptors, polymers, blends, and perovskites through a user-friendly interface. These features make RI$-$Calc a useful platform not only for studying materials (aiming at optoelectronic applications) but also as an auxiliary  tool for teaching purposes. Additionally, RI$-$Calc serves as a complementary program to our web server \href{https://nanocalc.org/}{nanocalc.org}, which is specifically geared towards organic optoelectronics applications.

\section*{Declaration of Competing Interest}
\noindent The authors declare that they have no known competing financial interests or personal relationships that could have appeared to influence the work reported in this paper.

\section*{Acknowledgments}

\noindent The authors acknowledge financial support from CNPq (grant 381113/2021$-$3), LCNano/SisNANO 2.0 (grant 442591/2019$-$5), INCT $-$ Carbon Nanomaterials and INCT $-$ Materials Informatics. L.B. (grant E$-$26/202.091/2022 process 277806), O.M. (grant E$-$26/200.729/2023 process 285493)  and G.C. (grant E$-$26/200.627/2022 and E$-$26/210.391/2022 process 271814) are greatfully for financial support from FAPERJ. The authors also acknowledge the computational support of N\'{u}cleo Avan\c{c}ado de Computa\c{c}\~{a}o de Alto Desempenho (NACAD/COPPE/UFRJ), Sistema Nacional de Processamento de Alto Desempenho (SINAPAD) and Centro Nacional de Processamento de Alto Desempenho em S\~{a}o Paulo (CENAPAD$-$SP) and technical support of SMMOL$-$solutions in functionalyzed materials.

\section{Data availability}
\noindent Data will be made available on request.

\bibliographystyle{elsarticle-num}
\bibliography{ri-bib}

\end{document}